\documentclass[conference]{IEEEtran}
\IEEEoverridecommandlockouts
\usepackage{cite}
\usepackage{amsmath,amssymb,amsfonts}
\usepackage{algorithmic}
\usepackage{graphicx}
\usepackage{textcomp}
\usepackage{xcolor}
\usepackage{balance}
\usepackage{multirow}

\usepackage{outlines}
\usepackage{physics}
\usepackage{tikz}
\usetikzlibrary{quantikz}
\usepackage{braket}
\usepackage[bf]{caption} 
\usepackage{adjustbox}
\usepackage{subcaption}

\usepackage{orcidlink}
\usepackage{amsmath}
\DeclareMathOperator*{\argmax}{arg\,max}
\def\BibTeX{{\rm B\kern-.05em{\sc i\kern-.025em b}\kern-.08em
    T\kern-.1667em\lower.7ex\hbox{E}\kern-.125emX}}
\begin{document}

\title{An Efficient Quantum Euclidean Similarity Algorithm for Worldwide Localization}

\author{Ahmed Shokry\orcidlink{0000-0003-3753-8886}, ~\IEEEmembership{Member, ~IEEE} and Moustafa Youssef\orcidlink{0000-0002-2063-4364}, ~\IEEEmembership{Fellow, ~IEEE}
\thanks{Ahmed Shokry is with the American University in Cairo, Egypt (e-mail: ahmed.shokry@aucegypt.edu).}
\thanks{Moustafa Youssef is with the American University in Cairo and the University of New South Wales, Australia (e-mail: moustafa-youssef@aucegypt.edu).}}

\maketitle

\begin{abstract}
Fingerprinting techniques are widely used for localization because of their accuracy, especially in the presence of wireless channel noise. However, the fingerprinting techniques require significant storage and running time, which is a  concern when implementing such systems on a global worldwide scale.

In this paper, we propose an efficient quantum Euclidean similarity algorithm for wireless localization systems. The proposed quantum algorithm offers exponentially improved complexity compared to its classical counterpart and even the state-of-the-art quantum localization systems, in terms of both storage space and running time. The basic idea is to entangle the test received signal strength (RSS) vector with the fingerprint vectors at different locations and perform the similarity calculation in parallel to all fingerprint locations. We give the details of how to construct the quantum fingerprint, how to encode the RSS measurements in quantum particles, and finally; present the quantum algorithm for calculating the euclidean similarity between the online RSS measurements and the fingerprint ones.

Implementation and evaluation of our algorithm in a real testbed using a real IBM quantum machine as well as a simulation for a larger testbed confirm its ability to correctly obtain the estimated location with an exponential enhancement in both time and space compared to the traditional classical fingerprinting techniques and the state-of-the-art quantum localization techniques. 

\end{abstract}


\section{Introduction}

Currently, location tracking services have become increasingly important, serving numerous applications in both indoor~\cite{youssef2015towards} and outdoor~\cite{shokry2018deeploc}. These include emergency services, navigation, location-based analytics, and many others. 

 Fingerprint-based localization techniques have a growing interest due to their accuracy. These techniques work in two phases. In the offline phase, the Radio Frequency (RF) fingerprint is built by scanning the received signal strength (RSS) signature of heard reference points (RPs\footnote{These reference points can be, e.g., WiFi access points; cellular cell towers; or Bluetooth beacons.}) at different locations in a database. Then in the online tracking phase, the set of overheard RPs is matched against fingerprints in the database to find the closest location in the RSS space to the unknown location. 
 
 Fingerprinting techniques share a fundamental characteristic, which is the requirement to compare online and offline measurements at each fingerprint location. As a result, their time and space complexity is $o(NM)$, where $N$ is the number of fingerprint locations and $M$ is the number of RPs. This approach is not scalable for supporting fingerprint-based indoor/outdoor localization on a global scale, especially for IoT environments, where the number of RPs with wireless interfaces in an environment can be significant.
Moreover, the localization accuracy of fingerprinting techniques is influenced by the fingerprint density (i.e., the number of fingerprint locations collected in the area, $N$) and the number of RPs used, $M$. Hence, a greater number of fingerprint locations and RPs result in more precise positioning. However, the time required to match the heard RSS with the fingerprint data also increases quadratically with the increase in the number of fingerprint locations and RPs.

To overcome the classical techniques' limitations, quantum fingerprint matching algorithms have been recently introduced ~\cite{quantum_arx, quantum_vision, quantum_qce, quantum_lcn, shokry2022device, shokry2023quantum}. In~\cite{quantum_vision, quantum_arx}, authors provide a general perspective on the use of quantum computing in location tracking and spatial systems. In~\cite{quantum_lcn}, they propose a cosine similarity quantum algorithm for localization systems. This work is further extended in ~\cite{shokry2022device, shokry2023quantum} to introduce a device-independent quantum fingerprint to enable heterogeneous devices localization. They also analyze the impact of noise and practical constraints on the intrinsic features of the quantum processors~\cite{quantum_qce}. 

These quantum localization techniques are built based on cosine similarity to attain sub-quadratic complexity, with a time and space complexity of $o(Nlog(M))$. In this paper, we present an efficient Euclidean distance quantum fingerprint matching algorithm that requires $o(\log(NM))$ space and runs in $o(\log(NM))$, i.e. sub-linear complexity, providing a promising technique that can scale to the large number of RPs and fingerprinting locations for worldwide next generation localization systems. This is more than exponentially better than its classical counterpart in both the number of RPs ($N$) and the size of the fingerprint ($M$). Moreover, it is more than exponentially better than the current
quantum localization algorithms in the fingerprint size ($M$). \footnote{
The proposed algorithm complexity of $o(log(NM))$ is \textbf{more than exponentially better} than the other state-of-the-art quantum algorithms complexity of $o(Nlog(M))$.}.

We validate our quantum localization algorithm on an instance of the IBM Quantum Experience real machine and discuss its performance. Our results shows that we can obtain the same  localization accuracy as the classical localization algorithms. This comes with exponential saving in time and space.

The rest of the paper is organized as follows: we start with a brief background on fingerprint-based localization in Section~\ref{sec:background}. Section~\ref{sec:qfp} provides the details of our quantum fingerprint matching algorithm. We then evaluate the proposed quantum algorithm in Section~\ref{sec:eval}. Finally, sections \ref{sec:related} and \ref{sec:conclude} discuss related work and conclude the paper, respectively.

\section{Background}
\label{sec:background}
\begin{figure}[!t]
	\centerline
	{\includegraphics[width=0.5\textwidth]{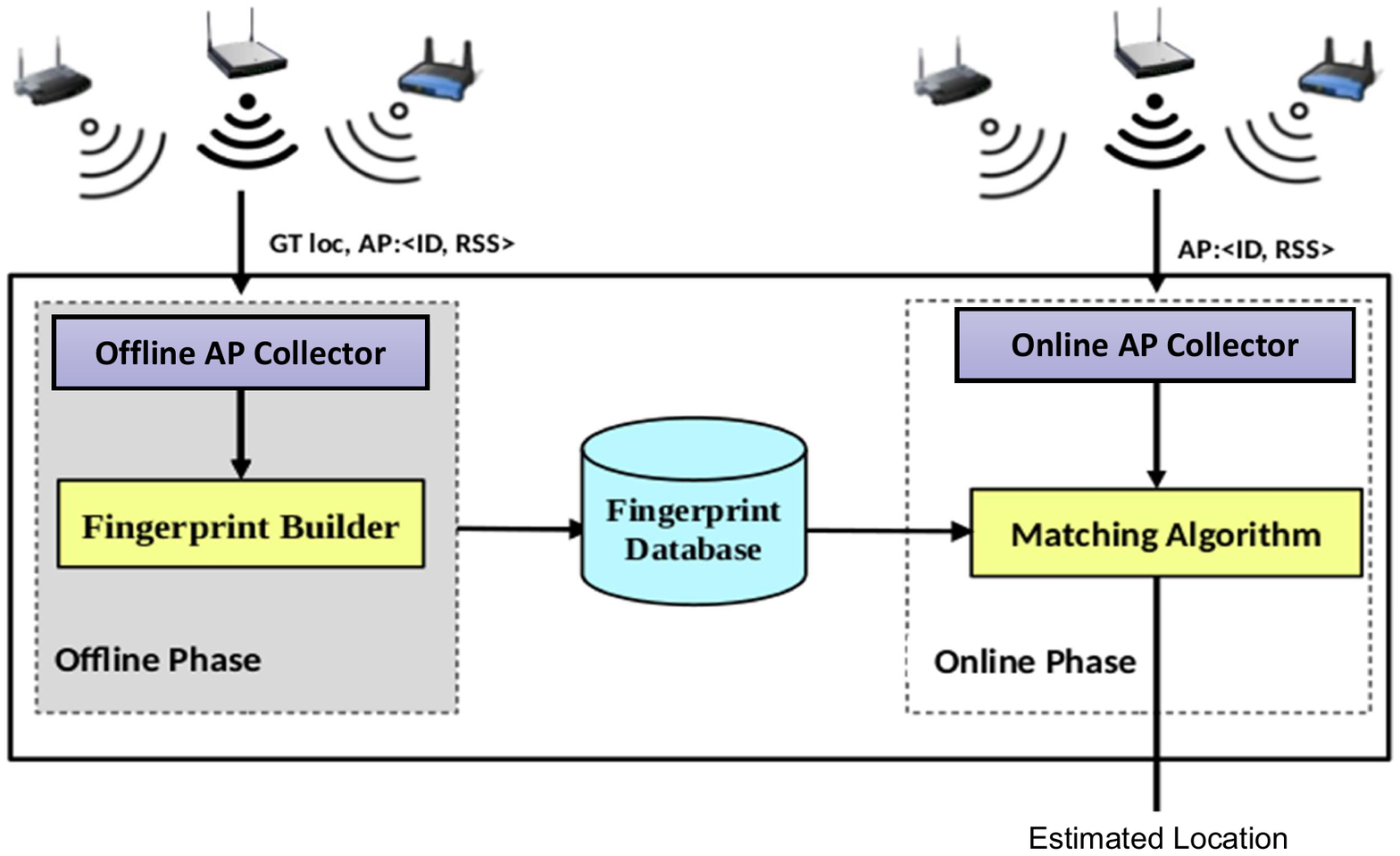}}
	\caption{General architecture for fingerprinting localization techniques~\cite{quantum_qce}.}
	\label{fig:arch}
\end{figure}

Fingerprint-based localization is a technique used to determine the location of a mobile device or a person by analyzing the unique fingerprint patterns of the Radio Frequency (RF) signals in the area of interest~\cite{bahl2000radar}. It relies on the fact that the signal strengths of the RPs, such as Wi-Fi or Bluetooth, vary depending on the distance and obstacles between the device and the RP. The overall framework for fingerprinting localization techniques is presented in Figure~\ref{fig:arch}. 

The process of fingerprinting localization includes two stages. The first stage is an offline phase. During this phase, the \textit{Offline AP Collector} collects the RSS measurements of the various RPs within the specific area of interest at various discrete ground-truth (GT) locations. The \textit{Fingerprint Builder} subsequently stores the RSS measurements vectors alongside their ground-truth locations in the fingerprint database. In the second stage, which is the online tracking phase, the \textit{Online AP Collector} scans the online RSS received from the various RPs at an unknown user location. The \textit{Matching Algorithm} then compares the online RSS vector with the fingerprint database. The matching algorithm identifies the location in the fingerprint that has the closest resemblance to the received signal, which is then regarded as the estimated location.
The fingerprinting techniques use different distance/similarity metrics to match the online RSS measurements with the ones stored at each location in the fingerprint. Examples include Euclidean, Manhattan, Chi-Squared, Bray-Curtis, and Mahalanobis, and cosine similarity~\cite{bahl2000radar,cos_sim1, cos_sim2,del2009efficient, beder2012fingerprinting}. In this paper, we use euclidean similarity, which is one of the earliest and most commonly used metrics in localization.

\section{The Proposed Quantum Localization Algorithm}
\label{sec:qfp}

In this section, we present the quantum version of the commonly used fingerprint-matching localization algorithms. 

We start the section by presenting the general quantum fingerprinting localization algorithm based on the Euclidean distance followed by a detailed example of how it works in a simple setting. We then 
discuss its efficiency and different aspects of its operation.

\begin{figure*}[t!]
	\centering
	\begin{adjustbox}{width=0.9\textwidth}
		\begin{tikzpicture}[row sep=0.1cm]
			\node at (-3.8,0.3) {Step A};
			\node at (-2.0,0.3) {Step B};
			\node at (0.5,0.3) {Step C};
			\node at (3.5,0.3) {Step D};
			\node at (5.0,0.3) {Step E};
			\draw[red,line width=1.6pt,dashed] (-4.6,0) -- (-4.6,-4.5);
			\draw[red,line width=1.6pt,dashed] (-3.2,0) -- (-3.2,-4.5);
			\draw[red,line width=1.6pt,dashed] (-0.9,0) -- (-0.9,-4.5);
			\draw[red,line width=1.6pt,dashed] (3.1,0) -- (3.1,-4.5);
			\draw[red,line width=1.6pt,dashed] (4.1,0) -- (4.1,-4.5);
			\draw[gray,line width=1.6pt,dotted] (-4.4,0) -- (2.9,0) -- (2.9,-4.4) -- (-4.4,-4.4) -- (-4.4,0);
			\node[text width=2.2cm] at (-7.0,-0.6) {Ancilla};
			\node[text width=2.4cm] at (-7.0,-1.7) {Index};
			\node[text width=2.4cm] at (-7.0,-2.8) {Data};
			\node[text width=2.6cm] at (-7.3,-3.9) {Location};
			\node[text width=4cm] at (-0.5,-4.7) {Quantum State Preparation};
			\node at (0,-0.1) [anchor=north]{
				\begin{quantikz}
			\lstick{$\ket{a} = \ket{0}$} & \qw & \gate{H} & \ctrl{2} & \gate{X} & \ctrl{2} & \qw & \qw & \gate{H} &\meter{} &   \cw \rstick{$p(a=1)$}\\
			\lstick{$\ket{i} = \ket{0}^{\otimes n}$}  & \qwbundle{n} & \gate{H^{\otimes n}} & \qw & \qw & \ctrl{1} & \gate{\text{++}} & \ctrl{2} & \qw & \qw & \qw & \qw \\	
			\lstick{$\ket{d} = \ket{0}^{\otimes m}$}  & \qwbundle{m} & \qw & \gate{\text{$R_\psi$}} & \qw & \gate{\text{$R_{\phi_i}$}} & \qw & \qw & \qw & \qw & \qw & \qw\\	
			\lstick{$\ket{l} = \ket{0}^{\otimes n}$}  & \qwbundle{n} & \qw & \qw & \qw & \qw & \qw & \gate{++} & \qw & \meter{} & \cw \rstick{$p(l_i|a=1)$}\\						
				\end{quantikz}
			};	
		\end{tikzpicture}
	\end{adjustbox}
	\caption{General quantum circuit for fingerprint-based localization based on Euclidean distance similarity.}
	\label{fig:comp_circ}
\end{figure*}
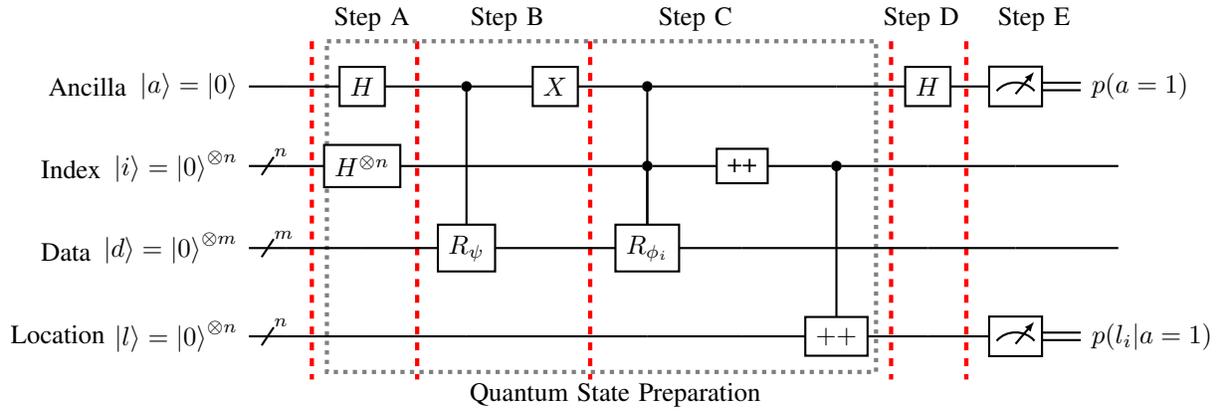

\subsection{General Quantum Localization Algorithm}
\label{sec:gen_alg}
Figure \ref{fig:comp_circ} shows the general quantum localization circuit. At a high level, the circuit encodes the online and the offline fingerprint RSS vectors into the amplitudes of the quantum system state, entangles them through quantum gates, and allows calculating the similarity to all fingerprints in parallel. This leads eventually to an exponential enhancement in both space and time as we analyze later. 

Without loss of generality, assume an area of interest with $N= 2^n$ discrete fingerprint locations and $M= 2^m$ RPs. The $N$ fingerprint location indices (from $0$ to $N-1$) can be encoded in $n$ qubits. Similarly, the M-dimensional normalized RSS vector from the $M$ RPs $\psi= (\alpha_0, \alpha_2, ..., \alpha_{M-1}), \sum_{i=0}^{M-1}\alpha_i^2 =1$, can be encoded using an $m$ qubits register (notice the exponential saving in state) $\ket{\psi}$ (or $\phi$ as $\ket{\phi} = \sum_{i=0}^{M-1} \beta_i \ket{i}, \sum_{i=0}^{M-1}$ $\beta_i^2 =1$), where the basis state $\ket{i}$ represents the binary encoding of integer $i$ \cite{qmemory,state_prep}. Assume the online RSS vector is $\ket{\psi}$ and the offline RSS vectors are $\ket{\phi_i}$, where $i$ is the fingerprint location index ($0 \leq i \leq N-1$) and the size of $\psi$ and $\phi_i$ is $M$
The general state preparation stage, i.e. mapping a classical vector to a quantum register can be achieved efficiently using different quantum circuits, e.g. \cite{qmemory,state_prep}. We give an example of how to prepare this state from classical vectors in the next section.

The input to the quantum circuit is a single Ancilla qubit $\ket{a}$ and three quantum registers with a total of $q=2n+m+1$ qubits. The first register is the index register $\ket{i}$ where we encode the index of each fingerprint vector. The second is the data register $\ket{d}$ where we encode the test RSS vector as well as the fingerprint RSS vectors. Finally, the location register where we encode the different location labels. Note that if we have one fingerprint vector at each location, then the size of the fingerprint index register is the same as the size of the location register. The output from the circuit is the probability distribution for each location in the fingerprint. Initially, the combined system state is 

\begin{equation}
\label{eq:initial}
    \ket{a\ i\ d\ l} = \ket{0}\ket{0}^{\otimes n}\ket{0}^{\otimes m}\ket{0}^{\otimes n}
\end{equation}
Where $n=\log(N)$ and $ m=\log(M)$.

First, the ancilla and the index qubits are put into a superposition state (Step A). Hence, the system state becomes
\begin{equation}
\frac{1}{\sqrt{2N}}\sum_{i=0}^{N-1} [\ket{0}\ket{i} + \ket{1}\ket{i} ] \ket{0}^{\otimes m}\ket{0}^{\otimes n}
\end{equation}

Then, the online test vector $\psi$ is entangled with the ground state of the ancilla (Step B)
\begin{equation}
\frac{1}{\sqrt{2N}}\sum_{i=0}^{N-1} [\ket{0}\ket{i} \ket{\psi}+ \ket{1}\ket{i} \ket{0}^{\otimes m}] \ket{0}^{\otimes n}
\end{equation}

Step C ends the state preparation phase where the different fingerprint vectors $\phi_i$ are encoded and entangled with the excited state of the ancilla. The (++) gate is a customized gate to encode and entangle each fingerprint data ($\ket{\phi_i}$) with the binary representation of its index in $\ket{i}$ and the binary representation of its location in $\ket{l}$. We can implement the (++) gate using NOT gates that take a quantum register at state $\ket{0}^{\otimes n}$ as input and keep increasing the index while entangling the fingerprint data corresponding to that index. We give a detailed example of how to implement (++) gate in Section~\ref{sub:ex}. 
Hence, the general state preparation stage takes the system to the following combined state
\begin{equation}
\frac{1}{\sqrt{2 N}} \sum_{i=1}^N \Big( \ket{0} \ket{i} \ket{\psi} + \ket{1} \ket{i} \ket{\phi_i} \Big) \ket{l_i} . 
\label{Eq:prep}
\end{equation}

This effectively creates an amplitude vector which contains the fingerprint vectors, $\ket{\phi_i}$, as well as $N$ copies of the online testing vector, $\ket{\psi}$. In step D, the Hadamard gate interferes with the copies of $\ket{\psi}$ with the fingerprint vectors $\ket{\phi_i}$, leading the system to the following general state
\begin{equation}
\frac{1}{2\sqrt{ N}} \sum_{i=0}^{N-1}  \Big( \ket{0} \ket{i} (\ket{\psi} + \ket{\phi_i} \big) +\ket{1} \ket{i} \big( \ket{\psi} - \ket{\phi_i} \big)  \Big) \ket{l_i}
\label{Eq:secH}
\end{equation} 

By normalizing the joint state
\begin{equation}
\begin{multlined}
\frac{1}{2\sqrt{ N}} \sum_{i=1}^N  
\Big( \sum_{i=0}^{N-1} |\ket{\psi}+\ket{\phi_i}|\ \ket{0}
\big[\ket{i} \frac{\ket{\psi} + \ket{\phi_i}}{\sum_{i=0}^{N-1} |\ket{\psi}+\ket{\phi_i}|} \big] \\ \ \ 
+\sum_{i=0}^{N-1} | \ket{\psi}-\ket{\phi_i}|\ \ket{1}\big[ \ket{i} \frac{\ket{\psi} - \ket{\phi_i}}{\sum_{i=0}^{N-1} |\ket{\psi}-\ket{\phi_i}|}  \big] \Big) 
\ket{l_i}
\end{multlined}
\end{equation} 

The probability that the ancilla qubit is in state $\ket{1}$ is
\begin{equation}
p(a=1) = \frac{1}{4N}  \sum_{i=1}^N |\ket{\psi} - \ket{\phi_i}|^2
\end{equation}

In Step E, the location register is measured conditioned on measuring the ancilla qubit to be in the $\ket{1}$ state.
This leaves the system in the following state
\begin{equation}
\frac{1}{2\sqrt{ N p_1}} \sum\limits_{i=1}^N \ket{i} \left( \ket{\psi} - \ket{\phi}_i\right)  \ket{l_i}
\end{equation}


This makes the amplitudes weight the location register $\ket{l_i}$ by the distance of the $i$th RSS vector in the fingerprint to the new online RSS vector. Hence, the probability of measuring a location qubits $l_i$ to be the estimated location is
\begin{equation}
\begin{multlined}
p(l_i)= \frac{1}{4 N p_1} \sum\limits_{i| l = l_i}  |\ket{\psi} - \ket{\phi_i} |^2 
\end{multlined}
\end{equation}
The previous equation reflects the probability of predicting location $l_i$ to be the estimated location for the online RSS vector. Therefore, the estimated location $l^*$ becomes
\begin{equation}
l^* = \argmax_{l_i} \ \  p(l_i)
\end{equation}

Practically, we repeat this circuit $K$ times to estimate the probability distribution for the fingerprint locations.

\subsection{Example}
\label{sub:ex}
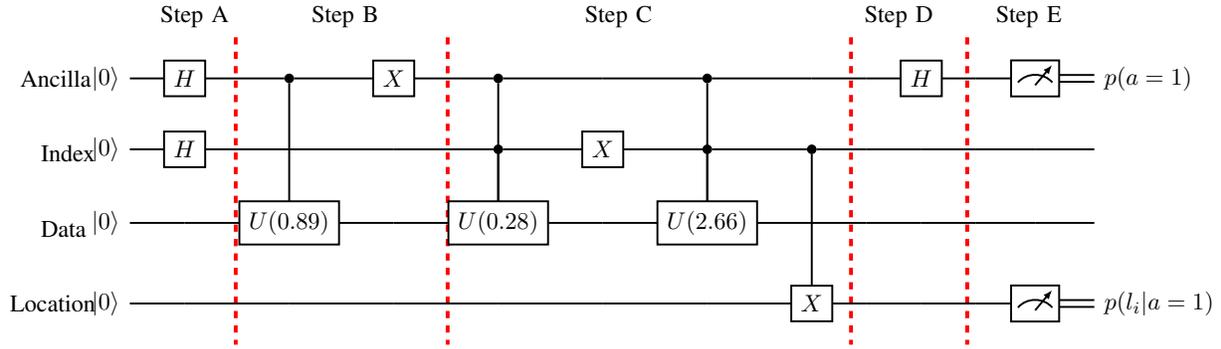
\begin{figure*}[t!]
	\centering
	\begin{adjustbox}{width=0.9\textwidth}
		\begin{tikzpicture}[row sep=0.1cm]
			\node at (-6.7,0.3) {Step A};
			\node at (-4.5,0.3) {Step B};
			\node at (-0.5,0.3) {Step C};
			\node at (3.6,0.3) {Step D};
			\node at (5.5,0.3) {Step E};
			\draw[red,line width=1.6pt,dashed] (-6.1,0) -- (-6.1,-4.5);
			\draw[red,line width=1.6pt,dashed] (-3.0,0) -- (-3.0,-4.5);
			\draw[red,line width=1.6pt,dashed] (2.9,0) -- (2.9,-4.5);
			\draw[red,line width=1.6pt,dashed] (4.6,0) -- (4.6,-4.5);
			\node[text width=1.9cm] at (-8.3,-0.6) {Ancilla};
			\node[text width=1.9cm] at (-8.0,-1.7) {Index};
			\node[text width=1.9cm] at (-8.0,-2.8) {Data};
			\node[text width=2.2cm] at (-8.3,-3.9) {Location};
			\node at (0,-0.1) [anchor=north]{
				\begin{quantikz}
			\lstick{$\ket{0}$}  & \gate{H} & \ctrl{2} & \gate{X} & \ctrl{2} & \qw & \ctrl{2} & \qw &  \qw & \gate{H} & \qw &\meter{} &   \cw \rstick{$p(a=1)$}\\
			\lstick{$\ket{0}$}   & \gate{H} & \qw & \qw & \ctrl{1} & \gate{X} & \ctrl{1} & \ctrl{2} & \qw & \qw & \qw & \qw & \qw \\	
			\lstick{$\ket{0}$}   & \qw & \gate{U(0.89)} & \qw & \gate{U(0.28)} & \qw & \gate{U(2.66)} & \qw & \qw & \qw & \qw & \qw & \qw\\	
			\lstick{$\ket{0}$}   & \qw & \qw & \qw & \qw & \qw & \qw & \gate{X} & \qw & \qw & \qw & \meter{} & \cw \rstick{$p(l_i|a=1)$}\\						
				\end{quantikz}
			};	
		\end{tikzpicture}
	\end{adjustbox}
	\caption{ A detailed example of the quantum fingerprint matching circuit using two RPs only. The circuit shows the state preparation stage, i.e. how to map the online RSS vector $\psi=(0.43, 0.9)$ and the fingerprint RSS vectors  $\phi_0=(0.99, 0.14)$ and $\phi_1=(0.24, 0.97)$ to the quantum states $\ket{\psi}=0.43 \ket{0}+ 0.9 \ket{1}$, $\ket{\phi_0}=0.99 \ket{0}+ 0.14 \ket{1}$, and $\ket{\phi_1}=0.24 \ket{0}+ 0.97 \ket{1}$, respectively, starting from $\ket{0}$. Firstly, the ancilla and index qubits are put into uniform superposition (step A) and the online RSS vector $\ket{\psi}$  is entangled with the ground state of the ancilla (step B). Then, the training vector $\ket{\phi_0}$ is entangled with the excited state of the ancilla and the ground state of the index qubit followed by entangling training vector $\phi_1$ with the excited state of the ancilla and the index qubit which completes the initial state preparation (step C). Next, the Hadamard gate interferes the copies of $\ket{\psi}$ with the training vectors (step D) and the ancilla is measured followed by a measurement of the location qubits when the ancilla was found to be in the $\ket{0}$ state (step E).}
	\label{fig:comp_ex}
\end{figure*}

In this section, we explain the quantum localization algorithm described in the previous section using a simple example with two RPs and two fingerprint locations. In particular, assume that the normalized testing RSS vector is $\psi=(0.43,0.9)$ and the two normalized fingerprint RSS vectors are $\phi_0=(0.99, 0.14)$ and $\phi_1=(0.24, 0.97)$. 

The algorithm calculates the distance between the online test measurement ($\psi$) and the fingerprint ones ($\phi_0$ and $\phi_1$) and the location with the highest probability is returned as the estimated location.  
Figure~\ref{fig:comp_ex} shows the complete circuit for location estimation.

The circuit starts by the state preparation stage, i.e. mapping the RSS vectors $\psi$, $\phi_0$ and $\phi_1$ to the quantum equivalent $\ket{\psi}=0.43 \ket{0}+ 0.9 \ket{1}$, $\ket{\phi_0}=0.99 \ket{0}+ 0.14 \ket{1}$, and $\ket{\phi_1}=0.24 \ket{0}+ 0.97 \ket{1}$, respectively. This is achieved by using the $U$ gate, which is represented as,
\begin{equation}
U(\theta) = \begin{bmatrix}
	\cos(\theta/2) & -\sin(\theta/2) \\
	\sin(\theta/2) & \cos(\theta/2) 
\end{bmatrix}
\end{equation}

Where $\theta$ is double the angle between $\ket{0}$ and a given normalized RSS vector. In particular, for a vector $ v=(a, b)$, $\theta$ should be set to $2\times\arctan(\frac{b}{a})$. 

The fingerprint matching part is the same as the one described in Figure~\ref{fig:comp_circ}. Since we have only two fingerprint locations and two RPs, the quantum registers contain only one qubit (i.e. $n=m=1$).

\subsection{Discussion}
\label{sec:disc}
In the offline training phase, the proposed quantum localization algorithm stores a fingerprint of size $N \times M$ in just $q = 1 + \log(N) + \log(M) + \log(N)$ qubits. This makes its space complexity $o(\log(NM))$ and hence, it achieves  exponential saving in space compared to the classical fingerprinting techniques which require $o(N M)$.

In the user tracking phase, the online RSS vectors are encoded using $R_\psi$ gate before being entangled with the fingerprint in the state preparation stage. The general state preparation stage can be achieved efficiently using different quantum  circuits, e.g.\cite{qmemory,state_prep}. A common way to achieve state preparation is to use quantum random access memory (QRAM) \cite{qmemory,rebentrost2014quantum, zhao2019quantum, state_prep} in which the binary representation of $\beta_i$ is loaded in parallel into a qubit register and conditional rotations are performed to encode the RSS measurements as amplitudes in quantum registers. Hence, using an efficient state preparation algorithm, the proposed quantum algorithm achieves $o(\log (NM))$ time bound. Thus, it provides an exponential speedup in time compared to the classical fingerprinting techniques.


\section{Implementation and Evaluation}
\label{sec:eval}
In this section, we implement the proposed quantum localization algorithm and evaluate its performance in a typical testbed on the real IBM quantum Santiago machine as well as IBM quantum machine simulator.

\subsection{Environment Setup}
\begin{figure}[!t]
	\centerline
	{\includegraphics[width=0.45\textwidth]{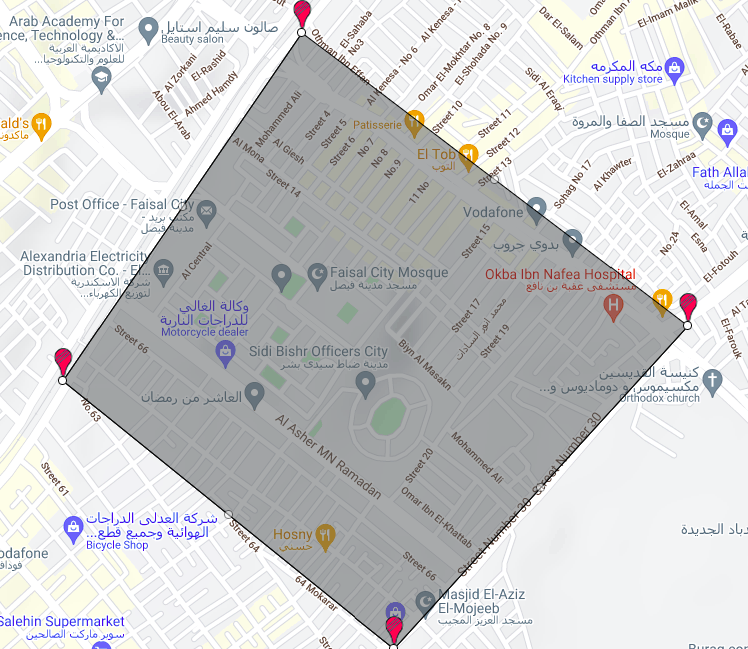}}
	\caption{The outdoor testbed.}
	\label{fig:testbed2}
\end{figure}

We evaluate the system in a typical outdoor cellular testbed (Figure~\ref{fig:testbed2}) that spans an 0.2Km$^2$ urban area. The area is covered by eight different cell towers. The fingerprint data is collected by war-driving at 16 discrete locations that uniformly cover the entire area of interest. We also collected an independent test set. Both the fingerprint data and test data locations are uniformly distributed over the entire area of interest. We use different Android devices for data collection. The deployed collector software collects GPS ground-truth locations, the cell towers received signal strengths, and timestamps.  

\subsection{Real Machine Experiment}
In the real machine experiment, we use the 5-qubits IBM Quantum Santiago machine. To fit the available number of qubits ($q = 2n+m+1 = 5$), we select only four cell-towers that are most commonly heard in the environment ($M=4$) and two fingerprint locations that are uniformly distributed across the environment ($N=2$). 

\begin{figure}[!t]
	\centerline
	{\includegraphics[width=0.5\textwidth]{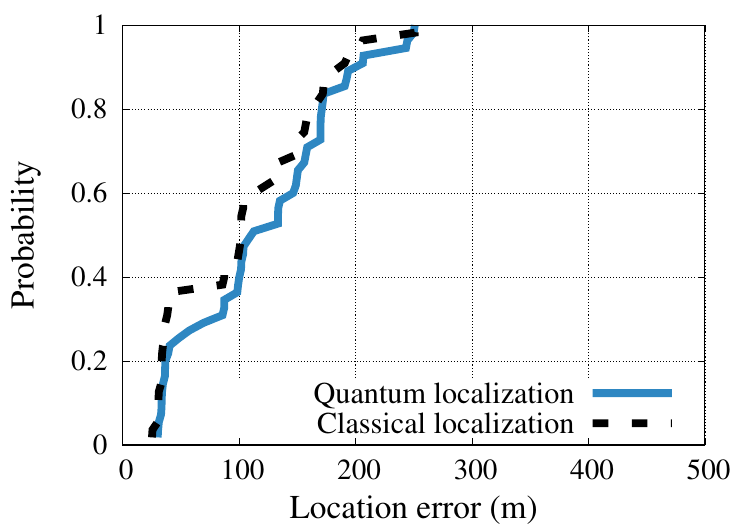}}
	\caption{CDF of localization error for the testbed using the IBM Quantum Santiago real machine.}
	\label{fig:rcdf}
\end{figure}

Figure \ref{fig:rcdf} shows the CDF of both the distance error for the quantum and the classical localization systems. The figure confirms that both systems have the same accuracy. This is achieved with the exponential gain compared to the classical system.

\subsection{Simulation Experiment} 
To show the performance  on a large scale testbed, we implemented the fingerprint matching circuit in Figure~\ref{fig:comp_circ} over the IBM Quantum machine simulator that has 28 qubits. The input to the quantum machine is a circuit diagram and the machine executes the circuit for several iterations/shots (parameter $K$ in Section~\ref{sec:gen_alg}) and returns the measurements results.

The eight cell towers in the outdoor testbed can be encoded using three qubits (note the logarithmic increase of the number of required qubits with the increase of the reference points). Therefore, we need a total of 12 qubits to run this experiment: one for the ancilla qubit, four qubits to model the index (which has a value between 0 and 15), four qubits for the class label for the 16 fingerprint locations, and three qubits to encode the fingerprint and testing data RSS which come from the eight cell towers in the environment.    

\begin{figure}[!t]
	\centerline
	{\includegraphics[width=0.5\textwidth]{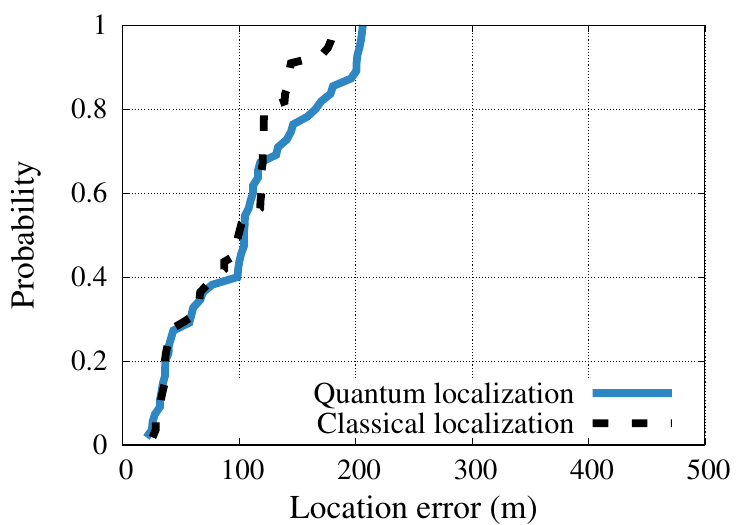}}
	\caption{CDF of localization error for the testbed using the IBM Quantum Simulator.}
	\label{fig:cdf}
\end{figure}

Figure \ref{fig:cdf} shows the CDF of distance error for the quantum and the classical localization systems. The figure confirms that the quantum fingerprinting localization system has the same performance as the classical one. This is achieved with the exponential potential gain of the quantum system in large scale as discussed in Section~\ref{sec:disc} and further confirmed in the next subsections.

\begin{figure}[t!]
	\centerline
	{\includegraphics[width=0.45\textwidth]{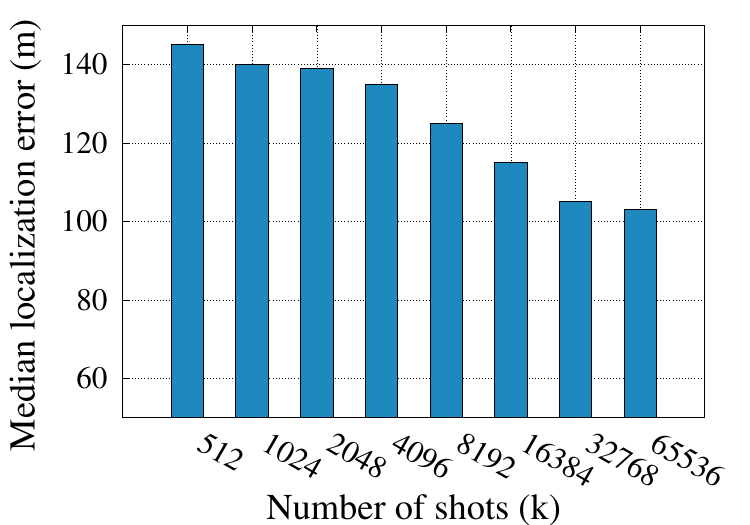}}
	\caption{Effect of changing number of shots (iterations) on the median localization accuracy.}
	\label{fig:shots}
\end{figure}

Figure \ref{fig:shots} shows the effect of increasing the number of shots, i.e. re-running the system (parameter $k$ in Section \ref{sec:gen_alg}), on the quantum system accuracy. The figure shows that, as expected, increasing the number of shots increases the system accuracy till it saturates around 32768 shots.

\begin{figure}[!t]
	\centerline
	{\includegraphics[width=0.5\textwidth]{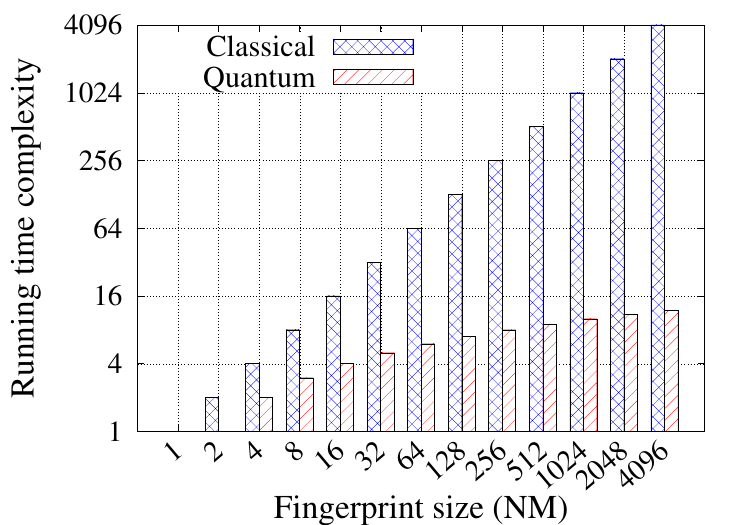}}
	\caption{Asymptotic running time.}
	\label{fig:time}
\end{figure}

Figure \ref{fig:time} further shows how the classical and quantum algorithms scale with the increase of the fingerprint size (number of fingerprint locations x number of RPs).  The figure confirms the exponential saving in running time of the proposed algorithm.

\section{Related Work}
\label{sec:related}
In this section, we discuss the classical and the quantum fingerprinting localization techniques. We also show how the proposed quantum localization algorithm differs from the existing work.

\subsection{Classical Fingerprinting-based Localization Systems}
Classical Fingerprinting-based localization consists of two main phases: offline training phase and online tracking phase~\cite{bahl2000radar, youssef2005horus}. During the offline training phase, a radio map of the area is created by collecting RF signal strength data at various known locations and storing them in a database. During the online tracking phase, the location of a mobile device is estimated by comparing its current RF signal strength data with the fingerprints stored in the database. 
There are different matching functions to compare the online RSS vector to those stored in the fingerprint. These matching functions include Euclidean, Manhattan, Chi-Squared, Bray-Curtis, Mahalanobis, and cosine similarity \cite{bahl2000radar,cos_sim1, cos_sim2,del2009efficient, beder2012fingerprinting}. The last one is usually used to combat the device heterogeneity effects \cite{cos_sim1, cos_sim2}. Similarly, probabilistic techniques, based on the MLE estimator, e.g. \cite{youssef2005horus}; have also been introduced.

The time and space complexity of the comparison process is $O(NM)$, where $N$ is the number of fingerprint locations and $M$ is the number of reference points (RPs). This means that as the number of fingerprint locations and RPs increases, the computational cost of the comparison process also increases, making the classical fingerprinting localization more time-consuming and resource-intensive.

\textit{On the other hand, the proposed quantum localization algorithm has time and space complexity of $O(log(NM))$ which is exponential enhancement in both time and space as we presented in this paper.}

\subsection{Quantum Fingerprinting-based Localization Systems}
Quantum algorithms have a promise for significant performance gains in different fields, including scientific computing \cite{moller2017impact}, robotics \cite{petschnigg2019quantum}, cryptography \cite{mavroeidis2018impact, cheng2017securing}, chemistry \cite{primas2013chemistry, improta2016quantum}, finance \cite{rebentrost2018quantum}, among others. To allow this, researchers have developed quantum algorithms to tackle general mathematical problems like solving linear systems of equations \cite{subacsi2019quantum}, linear differential equations \cite{xin2020quantum}, searching for an element in unsorted list \cite{grover}, and integer factorization \cite{shor}. 

Quantum algorithms substantially outperform their classical counterparts. For example, \cite{harrow2009quantum}, provides a quantum algorithm for solving linear systems of equations, which is exponentially faster than classical algorithms tackling the same problem. The well-known Grover algorithm searches for entries in an unsorted database of size $n$ in $O(\sqrt{n})$ steps \cite{grover}. The Shor algorithm provides a polynomial time quantum algorithm for integer factorization based on a quantum Fourier transform sub-module \cite{shor}.

Recently, quantum fingerprint matching algorithms have been introduced~\cite{quantum_arx, quantum_vision}. In~\cite{quantum_vision, quantum_arx}, the authors present a general overview of the use of quantum computing for localization and spatial systems. In~\cite{quantum_lcn, quantum_qce}, authors proposed a quantum cosine similarity-based localization algorithm. They extended the proposed quantum cosine similarity algorithm for heterogeneous devices localization in~\cite{shokry2022device, shokry2023quantum}. These techniques have time and space complexity of $O(Nlog(M))$ for $N$ fingerprint locations and $M$ RPs.

\textit{In this paper, we propose a fingerprint-based quantum algorithm for large-scale localization based on Euclidean similarity, which is one of the earliest and most commonly used metrics. The proposed technique achieves $O(log(MN))$ space and time complexity which is exponentially better than the state-of-the-art quantum localization algorithms.}

\section{Conclusion}
\label{sec:conclude}
We presented an efficient quantum euclidean similarity algorithm for worldwide localization. The proposed quantum algorithm achieves time and space complexity of $O(log(NM))$ which is exponentially better than the classical localization systems which need $O(NM)$, as well as the state-of-the-art quantum localization which needs $O(Nlog(M))$. We implemented the proposed algorithm on a real IBM Quantum machine as well as quantum simulator and evaluated it in a real testbed. The results confirm that we can achieve the same accuracy as the classical fingerprinting localization but with exponential saving in both the running time and the space complexity. This highlights the promise of our algorithm to scale fingerprinting localization systems to be used globally. 
   
\balance
\bibliographystyle{unsrt}
\bibliography{main.bbl}
\end{document}